\documentclass[showkeys,showpacs,prd,superscriptaddress]{revtex4}
\usepackage[dvips]{graphicx}
\usepackage{color}
\usepackage{amsmath}
\usepackage[dvips]{graphicx}
\begin{document}
\newcommand{\be}{\begin{equation}}
\newcommand{\ee}{\end{equation}}

\title{Local Approach to Hawking Radiation}

\author{Ari Peltola}
\email[Electronic address: ]{a.peltola-ra@uwinnipeg.ca} 
\affiliation{Department of Physics, The University of Winnipeg, 515 Portage Avenue, Winnipeg, Manitoba. Canada. R3B 2E9}

\begin{abstract} 
We consider an approach to the Hawking effect which is free of 
the asymptotic behavior of the metric or matter fields, and which
is not confined to one specific metric configuration. As a result,
we find that for a wide class of spacetime horizons there exists
an emission of particles out of the horizon. As expected, the energy 
distribution of the radiating particles turns out to be thermal.
  
\end{abstract}


\maketitle

There is a consensus between the researchers of general relativity 
that black holes emit thermal radiation with a spectrum similar to 
that of a black body. This peculiar phenomenon was first noticed by
Hawking when he considered the time evolution of the quantized real-valued 
scalar field in a spacetime produced by the collapse of a spherically
symmetric star \cite{haw}. Since then numerous papers have been 
written on this subject, and they confirm that all event horizons,
as well as some other known horizons spacetime (e.g. Rindler horizon), 
emit radiation \cite{path,pawi,unruh,bd}.

Although the analyses of the Hawking effect presented so far have
been very successful in the development of black hole thermodynamics,
there are still certain obstacles to cross. Firstly, many of the
approaches (including, for instance, Hawking's original work) are based
on the analysis of quantized matter at the asymptotic infinities. Hence
there is no straightforward way to generalize such approaches to 
situations where asymptotic infinities are absent. Secondly, the 
analyses presented so far are usually performed on one specific 
background metric (most often on Schwarzschild metric). However, 
if the Hawking effect is, as one believes, a phenomena which is present 
at \emph{any} horizon of spacetime, then there is certainly a need for a 
more general approach, powerful enough to provide the details of the 
radiation spectrum without explicit knowledge of the form of the metric.

In this paper, we consider an approach to the Hawking effect which
is both free of the asymptotic behavior of the metric or matter fields,
and, at the same time, not confined to one specific metric configuration. 
The motivation for this approach can be found from the findings of Ref. 
\cite{pad}. In that paper it was found that for a subclass of spacetimes
with a fixed value for their temperature, arising from the Euclidean
continuation of the spacetime metric, it is possible to define the concepts 
of energy and entropy. Curiously, all of the thermodynamical quantities were
defined in a way which depended only on the properties of the metric \emph{at 
the horizon}. In particular, the concept of energy was not determined by the
asymptotic flatness of the spacetime metric. As we shall soon see, in
those spacetimes there also exists, under certain restrictions, a thermal 
flux of particles coming out of the horizon. In other words, we shall
see that the temperature of those spacetimes can be interpreted by means
of radiative effects.

When developing a quantum field theoretic approach to the Hawking effect,
one must inevitably define some kind of a vacuum state in curved spacetime. 
In this paper it is suggested that in the vicinity of a spacetime horizon 
a natural choice for the vacuum state is a vacuum experienced by an observer 
in a free fall moving in the direction perpendicular to the horizon \cite{obs}. 
Such a choice is supported, for instance, by the similar roles played by an 
inertial observer in special relativity and a freely falling observer in 
general relativity. Quite recently, this program was successfully applied 
for both of the horizons of a Reissner-Nordstr\"om black hole \cite{pm}. In 
this paper, we shall find more evidence suggesting that such a choice for 
a vacuum is indeed in perfect harmony with the known properties of Hawking 
radiation.

As in Ref. \cite{pad}, the starting point of our analysis is the metric
\be \label{eq:ds} ds^2 = -f(r)\, dt^2+\frac{dr^2}{f(r)}+dL^2_\perp ,
\ee
where a (smooth) function $f(r)$ vanishes when $r=r_i$ ($i=1,\ldots ,n$) 
with $f'(r_i)$ remaining finite, and $dL^2_\perp$ is independent of the time
coordinate $t$. This form of the metric includes a wide class of spacetimes 
known to the literature. For instance, if one takes $dL^2_\perp$ to be the 
metric of a 2-sphere and $r\in [0,\infty[$ to be the radial coordinate, 
Eq. (\ref{eq:ds}) covers spherically symmetric spacetimes such as the Schwarzschild 
spacetime, the Reissner-Nordstr\"om spacetime, and the de Sitter spacetime. If, on the 
other hand, one interprets $r$ as a Cartesian coordinate $x\in ]-\infty, \infty[$
and takes $dL^2_\perp = dy^2+dz^2$, Eq. (\ref{eq:ds}) includes Rindler spacetime.

As a first step, let us specify the different types of spacetime horizons
arising from the metric (\ref{eq:ds}) \cite{hor}. Consider an arbitrary coordinate 
singularity $r=r_i$. If $f'(r_i) \ne 0$, the function $f$ changes its sign at 
the surface $r=r_i$. Hence there is a static spacetime region on one side of 
that surface and, correspondingly, a non-static spacetime region on the opposite 
side of the surface. If $f'(r_i)> 0$, the static region is located on that side 
where $r >r_i$, whereas if $f'(r_i)< 0$, the static region is located on that side 
where $r <r_i$. To simplify the terminology, these static regions are referred 
here as the static regions located right or left from the horizon, respectively.
This terminology is consistent with the typical choice where the coordinate $r$ 
increases towards the right hand side of the $tr$-plane. More pathological cases 
may arise if $f'(r_i)=0$. If this happens, the behavior of the function $f$ in
the vicinity of $r=r_i$ is determined by its higher order derivatives $f^{(k)}(r_i)$
($k=2,3,\ldots$). There are cases where $f$ indeed changes its sign, but if $f(r_i)$
is a local minimum or maximum, then the surface $r=r_i$ separates only two static 
regions or two non-static regions, respectively. For every case where there is
a static region on either side of the surface $r=r_i$, it is possible to apply 
the approach given in this paper. However, it will turn out that a vanishing first
derivative at $r=r_i$ leads also to a vanishing temperature for the radiation. 
Therefore, the cases where $f'(r_i)=0$ are either trivial (i.e, there is no
radiation because the temperature is zero) or they lack physical significance 
(i.e., there are no static regions on either side of the surface $r=r_i$). 
From this point on, we shall concentrate only on the situations where 
$f'(r_i) \ne 0$.

To begin with this analysis, let us first study the spherical symmetric 
spacetimes where 
\be dL^2_{\perp}=r^2\big( d\theta^2+\sin^2\theta\, d\varphi^2\big)
\ee
and $r$ is the radial coordinate, and leave the planar symmetric spacetimes aside 
for a moment. For simplicity, we shall consider the situations where $f'(r_i)>0$ 
and $f'(r_i)<0$ separately. At first, let us take $f'(r_i)>0$ \cite{note}. In that case,
the static region is located right from the horizon $r=r_i$. The idea of this 
paper is to develop a quantum field theoretic approach which is valid at the 
immediate vicinity of the horizon---the region responsible for the Hawking effect.
To that end, consider the Klein-Gordon equation for massless particles,
\be \label{eq:KGE} g^{\mu \nu} D_\mu D_\nu \phi=0,
\ee
where $D_\mu$ denotes the covariant derivative. To further simplify this equation,
we note that the spacetime region near a horizon is effectively two-dimensional:
the transverse degrees of freedom are simply redshifted away relatively to the
ones in the $rt$-plane \cite{car}. Therefore, near the horizon one can write, as
an excellent approximation,
\be \label{eq:approx} \phi= \phi(t,r).
\ee
Using this simplification, and defining a new function $R(t,r)$ by
\be \label{eq:R} \phi(t,r) =: \frac{1}{r}\, R(t,r)
\ee
and the ``tortoise coordinate" $r_*$ by
\be \label{eq:tortoise} \frac{dr_*}{dr} := \frac{1}{f},
\ee
one finds that very close to the horizon, Eq. (\ref{eq:KGE}) reduces to
\be \label{eq:KGE1}\bigg[ \frac{\partial^2}{\partial t^2} -\frac{\partial^2}{\partial r^2_*}
  + V(r) \bigg] R(t,r) =0,
\ee
where the ``potential" $V(r)$ has the property
\be \label{eq:V} V(r) := \frac{f'f}{r} \xrightarrow{r\rightarrow r_i} 0.
\ee

It is now easy to solve the Klein-Gordon equation at the region close to the
horizon. It follows from Eqs. (\ref{eq:R}), (\ref{eq:KGE1}), and (\ref{eq:V}) 
that the orthonormal solutions are of the form
{\setlength\arraycolsep{2pt}
\begin{subequations}
\begin{eqnarray} \label{eq:sol1} \phi_\text{in} &=& Nr^{-1}e^{-i\omega V},\\ 
  \label{eq:sol2} \phi_\text{out} &=& Nr^{-1}e^{-i\omega U},
\end{eqnarray}
\end{subequations}}%
where $N$ is an appropriate normalization constant and the advanced and 
retarded coordinates $V$ and $U$ are defined as
{\setlength\arraycolsep{2pt}
\begin{subequations}
\begin{eqnarray} V &:=& t+r_*, \\ U &:=& t-r_*.
\end{eqnarray}
\end{subequations}}%
The solutions $\phi_\text{in}$ and $\phi_\text{out}$ represent, from the point
of view of an observer at rest with respect to the horizon, particles with ``energy" 
$\omega$ going in and coming out of the horizon, respectively. To be quite
precise, however, one should note that the quantity $\omega$ represents the 
energy of a particle when it is measured with respect to the time coordinate 
$t$. This means that, according to the observer at rest close the horizon, the 
quantity $\omega$ is not exactly the energy of a particle: it does not include 
the redshift resulting from the spacetime metric. Nevertheless, because $\omega$ is
so closely related to the energy of a particle (from the point of view of the
given observer), we shall continue to refer to this quantity simply as ``energy". 
The effects of redshift will be taken account later when we have found the 
effective temperature of the radiation. From this point on, we shall take 
$\omega >0$.

The next task is to consider Eq. (\ref{eq:KGE}) in the rest frame of a freely
falling observer traveling across the horizon. Let us first define a new set
of coordinates in the static region such that
{\setlength\arraycolsep{2pt}
\begin{subequations} \label{eqs:uv}
\begin{eqnarray} u &:=& \frac{1}{2}\big( e^{\kappa V} + e^{-\kappa U} \big) ,\\
v &:=& \frac{1}{2}\big( e^{\kappa V} - e^{-\kappa U} \big) ,
\end{eqnarray}
\end{subequations}}%
where
\be \kappa = \frac{f'(r_i)}{2}.
\ee
As the notation suggests, $u$ and $v$ are sort of ``generalized Kruskal coordinates", 
while $\kappa$ may be interpreted as the surface gravity at the horizon. Using these 
coordinates, the spacetime metric reads
\be \label{eq:ds2} ds^2 = \frac{f}{\kappa^2}\, e^{-2\kappa r_*}\big( -dv^2 + du^2\big) 
+ r^2\big( d\theta^2+\sin^2\theta\, d\varphi^2\big) .
\ee
It is easy to see that this metric is regular at the horizon. It follows from
the definition (\ref{eq:tortoise}) that close to the horizon
\begin{eqnarray} \label{eq:expansion} e^{-2\kappa r_*} &=& \exp \bigg( -f'(r_i) 
\int \frac{dr}{f'(r_i) (r-r_i)+\frac{1}{2}f''(r_i)(r-r_i)^2 +\cdots}\bigg) \nonumber \\
&=& \frac{1}{|r-r_i|}\, \exp \big[ Cr+ D +\mathcal{O} \big( (r-r_i)^2\big) \big],
\end{eqnarray}
where $C:=\frac{1}{2}\frac{f''(r_i)}{f'(r_i)}$ and the integration constant D can be 
determined by the requirement $r_* \xrightarrow{r\rightarrow 0} 0$. Hence the metric 
(\ref{eq:ds2}) is regular at the horizon.

The generalized Kruskal coordinates $u$ and $v$ can utilized when defining a geodesic 
coordinate system at the horizon. It follows from the construction given above that
the location of the (future) horizon of an observer at rest with respect to the 
spherical coordinates $r$, $\theta$, and $\varphi$ is given by the condition $u=v$.
Let us now choose a point $u=v=0$ from the horizon \cite{bif}, and define a new set
of coordinates such that
{\setlength\arraycolsep{2pt}
\begin{subequations} \label{eqs:geod}
\begin{eqnarray} X^0 &:=& \alpha v ,\\ X^1 &:=& \alpha u,
\end{eqnarray}
\end{subequations}}%
where
\be \label{eq:alfa}\alpha := \lim_{r\rightarrow r_i}\bigg( \frac{\sqrt{f}}{|\kappa |}\, 
e^{-\kappa r_*} \bigg).
\ee
Using these definitions, the spacetime metric at the given point takes the form
\be \label{eq:ds3} ds^2 = -\big( dX^0 \big)^2 + \big( dX^1 \big)^2 
+ r^2\big( d\theta^2+\sin^2\theta\, d\varphi^2\big)
\ee
It is easy to see that in this new system of coordinates a certain freely falling 
observer is momentarily at rest when $u=v=0$. For an observer moving 
in the direction perpendicular to the horizon one has $d\theta = d\varphi =0$, so 
it is sufficient to show that the coordinates $X^0$ and $X^1$ constitute a geodesic 
system of coordinates. This condition, in turn, is satisfied if the derivatives 
of the metric components $g_{00}$ and $g_{11}$ of Eq. (\ref{eq:ds3}) vanish when 
$u=v=0$. We shall now show that this is indeed the case. 

To begin with, we first note that the relationship between $X^0$ and $X^1$ can be 
written in the implicit form:
\be \label{eq:implicit} \big( X^1 \big)^2 - \big( X^0 \big)^2 = \alpha^2 e^{2\kappa r_*}.
\ee
By differentiating the both sides of this equation (separately) with respect to
$X^0$ and $X^1$, one obtains, respectively,
{\setlength\arraycolsep{2pt}
\begin{subequations} \label{eqs:Xder}
\begin{eqnarray} -X^0 &=& \alpha^2\kappa e^{2\kappa r_*}\frac{dr_*}{dr}\frac{\partial r}{\partial X^0},
	\\ X^1 &=& \alpha^2\kappa e^{2\kappa r_*}\frac{dr_*}{dr}\frac{\partial r}{\partial X^1}.
\end{eqnarray}
\end{subequations}}%
But when the coordinate $r_*$ is expanded as Taylor series similarly as in Eq. 
(\ref{eq:expansion}), one obtains
\be e^{2\kappa r_*}\frac{dr_*}{dr} \ne 0.
\ee
Hence
\be \frac{\partial r}{\partial X^0} = \frac{\partial r}{\partial X^1} = 0
\ee 
at the point where $u=v=0$. Because, according to Eqs. (\ref{eq:ds2}) and 
(\ref{eqs:geod}), the metric components $g_{00}$ and $g_{11}$ of Eq. (\ref{eq:ds3}) 
depend only on the coordinate $r$, one concludes that $X^0$ and $X^1$ constitute
a geodesic system of coordinates when $u=v=0$.

It is now time to write down the massless Klein-Gordon equation in a rest frame of 
the freely falling observer. For consistency with Eq. (\ref{eq:approx}), we shall
search for solutions of the form
\be \phi= \phi(X^0,X^1).
\ee
Using the metric (\ref{eq:ds3}) and defining a new function $\widetilde{R}(X^0,X^1)$ by
\be \label{eq:R2} \phi(X^0,X^1) =: \frac{1}{r}\, \widetilde{R}(X^0,X^1), 
\ee
Eq. (\ref{eq:KGE}) can be finally written as
\be \label{eq:KGE3}\bigg[ \frac{\partial^2}{\partial (X^0)^2} -\frac{\partial^2}{\partial (X^1)^2}
  + \widetilde{V}(r) \bigg] R(X^0,X^1) =0,
\ee
where the ``potential" $\widetilde{V}(r)$ is given by the definition:
\be \widetilde{V}(r) := \frac{1}{r}\bigg[ \frac{\partial^2 r}{\partial (X^1)^2} 
	- \frac{\partial^2 r}{\partial (X^0)^2}\bigg].
\ee
With help of Eqs. (\ref{eq:implicit}) and (\ref{eqs:Xder}), one finds that this 
potential has the property
\be \label{eq:tildeV} \widetilde{V}(r) = \frac{1}{r}\frac{ff'}{\alpha^2 \kappa^2}\, e^{-2\kappa r_*} 
	\xrightarrow{r \rightarrow r_i} \frac{f'(r_i)}{r_i}.
\ee
So it turns out that, in contrast to the potential $V(r)$ in Eq. (\ref{eq:V}), the
potential $\widetilde{V}(r)$ does not vanish at the horizon.

The question whether the potential $\widetilde{V}(r)$ can be ignored when obtaining
the solutions for Eq. (\ref{eq:KGE3}) is a very subtle one and requires much attention.
The first observation is that $\widetilde{V}(r)$ has an inverse dependency on the
curvature radius $r_i$ of the horizon, which suggests that in many situations this
potential becomes negligibly small. Indeed, if one considers macroscopic horizons 
alone, which means that in natural units $r_i \gg 1$, it turns out that $\widetilde{V}(r)$ 
becomes negligible for every single spherically symmetric horizon known to the literature
(examples of this will be given later). Therefore, for any physically relevant 
macroscopic horizon the effects of this potential may be ignored. There are, however, 
many physically interesting situations, where such approximation is certainly \emph{not} 
justified. Of special interest are the final stages of black hole evaporation: When 
the mass of a black hole becomes small enough, one can no more disregard the effects
of the potential $\widetilde{V}(r)$. In that case, one should rather solve Eq. (\ref{eq:KGE3}) 
by treating $\widetilde{V}(r)$ as a constant given by Eq. (\ref{eq:tildeV}) \cite{car2}.
The resulting differential equation would again yield
spherical wave solutions, but now with a constraint between energy $\omega$, 
wavenumber $k$, and the potential $\widetilde{V}(r)$. These solutions should lead to a 
radiation spectrum which may be radically different from the thermal spectrum of 
the Hawking radiation, but which is hardly analytically solvable. Actually, this 
kind of behavior is something that one might expect. For that reason, we strongly
endorse the idea that the potential $\widetilde{V}(r)$ may be of vital importance
when studying the final stages of black hole evaporation. As interesting this proposal
may be, in this paper we shall still concentrate on the macroscopic horizons and assume
that the potential $\widetilde{V}(r)$ is small enough to be neglected. However, the 
possible importance of this potential in the study of black hole evaporation should be 
fully examined in the future research.

When the potential of Eq. (\ref{eq:tildeV}) is small enough to be neglected, 
it follows from Eqs. (\ref{eq:R2}) and
(\ref{eq:KGE3}) that the solutions to the Klein-Gordon equation are of the form
{\setlength\arraycolsep{2pt}
\begin{subequations} \label{eqs:solutions}
\begin{eqnarray} \label{eq:sol3} \phi'_\text{in} &=& Nr^{-1}e^{-i\omega \widetilde{v}},\\ 
  \label{eq:sol4} \phi'_\text{out} &=& Nr^{-1}e^{-i\omega \widetilde{u}},
\end{eqnarray}
\end{subequations}}%
where
{\setlength\arraycolsep{2pt}
\begin{subequations}
\begin{eqnarray} \widetilde{v} &=& X^0+X^1, \\ \widetilde{u} &=& X^0-X^1.   
\end{eqnarray}
\end{subequations}}%
The solutions $\phi'_\text{in}$ and $\phi'_\text{out}$ represent, from the point 
of view of the freely falling observer, particles with energy $\omega>0$ going 
in and coming out of the horizon, respectively.

The next logical step is to ask whether the vacuum associated with the freely 
falling observer is different from the vacuum associated with the observer 
at rest with the respect to the horizon. The answer for this question is
provided by the so-called Bogolubov transformation \cite{bog}. Because the
relationship between the coordinates $U$ and $\widetilde{u}$ is given by 
the formula
\be U = -\kappa^{-1} \ln (-\widetilde{u}) + \kappa^{-1}\ln \alpha,
\ee
the Bogolubov transformation between the outcoming solutions (\ref{eq:sol2}) 
and (\ref{eq:sol4}) reads
\be e^{i\omega \kappa^{-1} \ln (-\widetilde{u}) -i\omega \kappa^{-1}\ln \alpha}=
	\sum_{\omega'} \big( A'_{\omega \omega'}e^{-i\omega'\widetilde{u}}+
	B'_{\omega \omega'}e^{i\omega'\widetilde{u}}\big),
\ee
where the Bogolubov coefficients $A'_{\omega \omega'}$ and $B'_{\omega \omega'}$ 
are expressible as Fourier integrals:
{\setlength\arraycolsep{2pt}
\begin{subequations}
\begin{eqnarray} A'_{\omega \omega'}&=& \frac{1}{2\pi}\, e^{-i\omega \kappa^{-1}\ln \alpha} 
	\int\limits_{-\infty}^0 d\widetilde{u}\, e^{i\omega \kappa^{-1} \ln (-\widetilde{u})}
	e^{i\omega' \widetilde{u}} \\ B'_{\omega \omega'}&=& \frac{1}{2\pi}\, e^{-i\omega \kappa^{-1}\ln \alpha} 
	\int\limits_{-\infty}^0 d\widetilde{u}\, e^{i\omega \kappa^{-1} \ln (-\widetilde{u})}
	e^{-i\omega' \widetilde{u}}. 
\end{eqnarray}
\end{subequations}}%
The integration is performed here from the negative infinity to zero 
because in the static region $\widetilde{u}<0$. The integrals given
above are similar to those of Refs. \cite{haw} and \cite{pm}, and 
together these integrals imply
\be |A'_{\omega \omega'}| = e^{\pi\kappa^{-1}\omega}|B'_{\omega \omega'}|.
\ee

One can now easily obtain the energy distribution of the radiating 
particles. It is well known that between the Bogolubov coefficients 
there is a relationship:
\be \sum_{\omega'} \big( |A'_{\omega \omega'}|^2 - |B'_{\omega \omega'}|^2\big)=1.
\ee
So one finds that when the field is in vacuum from the point of view 
of the freely falling observer, the number of the of particles coming 
out of the horizon with energy $\omega$ is, from the point of view of 
the observer at rest very close to the horizon,
\be n_\omega = \sum_{\omega'}|B'_{\omega\omega'}|^2 = \frac{1}{e^{2\pi\kappa^{-1}\omega}-1}.
\ee
This is a Planck distribution at the temperature
\be T_0 = \frac{\kappa}{2\pi} = \frac{f'(r_i)}{4\pi},
\ee
which represents the temperature of the radiation when the redshift
effects are ignored. This temperature is related to the actual 
temperature experienced by the observer at rest very close to the
horizon by the Tolman relation \cite{landau}:
\be \label{eq:T1} T_+ = (g_{00})^{-\frac{1}{2}}\, T_0 = \frac{1}{\sqrt{f(r)}}\frac{f'(r_i)}{4\pi}.
\ee

The equation given above applies for spherically symmetric horizons
with $f'(r_i)>0$. Let us now briefly discuss what kind of modifications 
take place when $f'(r_i)<0$. In this situation the static region is 
located left from the horizon, which means that the roles of the
solutions $\phi_{\text{in}}$ and $\phi_{\text{out}}$ interchange.
Hence the outcoming solutions to the Klein-Gordon equation behave, 
from the point of view of an observer at rest very close to the 
horizon, as
\be \label{eq:sol5} \phi_\text{out} = Nr^{-1}e^{-i\omega V}.
\ee
As it comes to the freely falling observer, one can still construct 
a rest frame for that observer through Eqs. (\ref{eqs:uv})-(\ref{eq:alfa}),
but it turns out that the coordinate $v$ becomes a decreasing function of
the time coordinate $t$ whereas the coordinate $u$ becomes a decreasing
function of the coordinate $r_*$. This means that, assuming the potential
(\ref{eq:tildeV}) is small enough to be neglected, the outcoming solutions 
from the point of view of the freely falling observer read
\be \label{eq:sol6} \phi'_\text{out} = Nr^{-1}e^{i\omega\widetilde{v}}.
\ee
The Bogolubov transformation between the outcoming solutions (\ref{eq:sol5}) 
and (\ref{eq:sol6}) implies that the effective temperature experienced
by the observer at rest very close to the horizon is
\be \label{eq:T2} T_- = \frac{1}{\sqrt{f(r)}}\frac{-f'(r_i)}{4\pi}
\ee
Taken as a whole, Eqs. (\ref{eq:T1}) and (\ref{eq:T2}) finally lead to
the temperature
\be \label{eq:T3} T = \frac{1}{\sqrt{f(r)}}\frac{|f'(r_i)|}{4\pi}
\ee
for the radiation of a spherically symmetric horizon. This result is in a 
perfect harmony with the findings of Ref. \cite{pad}.

A familiar example of a spherically symmetric spacetime is the Schwarzschild
spacetime with $f(r)=1-2M/r$, which has a coordinate singularity $r_\text{S}=2M$. 
In that case the potential (\ref{eq:tildeV}) behaves as $\widetilde{V} \sim 1/M^{2}$ 
which indicates that for all macroscopic Schwarzschild black holes, i.e., 
Schwarzschild black holes with mass well above the Planck mass, the potential 
$\widetilde{V}$ will indeed vanish at the horizon. The approximation used in 
Eqs. (\ref{eqs:solutions}) is therefore justified, and 
using Eq. (\ref{eq:T3}) one obtains the effective temperature
\be T_\text{S} = \frac{1}{8\pi M\sqrt{1-\frac{2M}{r}}}
\ee
for the black hole radiation, as measured by an observer at rest very close
to the horizon. The factor involving a square root is due to the redshift
effects close to the horizon. Another important example is the de Sitter 
spacetime, where $f(r)=1-H^2 r^2$ and $H$ is the Hubble constant. This metric
has a coordinate singularity $r_\text{dS}=H^{-1}$, and at the cosmological 
scales of distances the potential $\widetilde{V}$ will certainly vanish at the horizon. 
Hence the temperature of the radiation very close to the horizon has the form
\be T_\text{dS} = \frac{2H}{\sqrt{1-H^2r^2}}, 
\ee
where, again, the square root in the denominator is due to the redshift. 
According to the best of knowledge of the author, this is the first 
explicit derivation of this temperature by means of quantum field 
theoretic arguments. 

Let us next draw our attention to the plane symmetric spacetimes, where 
$r$ is interpreted as a Cartesian coordinate $x$ and 
\be dL^2_{\perp}= dy^2 + dz^2.
\ee
In essence, the analysis of the Klein-Gordon field close to a plane
symmetric horizon may be performed in a very similar way as in the 
spherical symmetric case. The only difference is that the Klein-Gordon 
equation no more includes a potential analogous to the potential 
$\widetilde{V}$ of Eq. (\ref{eq:tildeV}). The reason for this is easy
to understand. In the case of spherical symmetric horizons, it is
natural to expect that the properties of the radiation depend, in a
way or another, on the ``size" of the horizon. As we have seen, such
a dependency is given by the potential $\widetilde{V}$. The plane 
symmetric horizons, in turn, are not compact but infinite. Consequently,
there is no potential analogous to $\widetilde{V}$. The Bogolubov transformation 
between the outcoming solutions leads again to a Planck distribution, 
and taking account the redshift effects at the horizon, one finally
obtains the same temperature as was found in Eq. (\ref{eq:T3}).

To see that Eq. (\ref{eq:T3}) really holds in a plane symmetric case, 
consider, as an example, the Rindler spacetime. By choosing $f=2x-1$ 
and $x=(1/a^2+1)/2$, where $a$ is the proper acceleration, the metric 
(\ref{eq:ds}) describes the Rindler spacetime. Equation (\ref{eq:T3}) 
then implies
\be \label{eq:unruh} T = \frac{a}{2\pi}.
\ee
Note that here the factor $a$ arises from the redshift.

Although Eq. (\ref{eq:unruh}), the Unruh temperature, is a well-known 
result, the treatment given above has still some significance on the 
research of today. Of special interest are the objects often referred
as ``local Rindler horizons" \cite{jac}. In broad terms, a local Rindler
horizon is a horizon which appears in the rest frame of a uniformly
accelerated observer---even when the spacetime is curved. When the
curvature of spacetime is reasonably small (which usually means that
the curvature of spacetime is not significant at the Planck scale of 
distances \cite{mes}), an accelerated observer located very close to 
his local Rindler horizon will see his surroundings as a piece of 
the Rindler spacetime. Because of that the analysis given here
may be applied to the local Rindler horizons as well. This is an important 
observation since the local Rindler horizons are physically more 
realistic than the Rindler horizons in a flat spacetime. Even better, 
such observation opens up a possibility to study the radiation of 
asymmetric horizons since from a close distance most of the horizons 
appear similar to a Rindler horizon. This is a strong argument in 
favor of a (still controversial) proposal that all horizons of
spacetime radiate.

The examples given above show that Eq. (\ref{eq:T3}), which was
derived here by means of quantum field theoretic arguments, is in
a perfect harmony with the known properties of Hawking radiation.
However, because the function $f$ is arbitrary, this result is 
\emph{not} confined to the known solutions of Einstein's field
equation but the similar chain of reasoning may be applied to more
general forms of the metric. Indeed, since Einstein's field equation
was nowhere used in this paper, one arrives at the conclusion that
the Hawking effect is \emph{not} a product of Einstein's equation, but
instead it is a pure consequence of the existence of a horizon.
The obvious advantage of the approach considered 
here is that the Hawking effect is treated as a local phenomenon near
a horizon. In fact, even though the metric (\ref{eq:ds}) seems to 
suggests the existence of a global Killing field, such an assumption
is by no means essential or even necessary: As the analysis is performed 
at the local neighborhood of the horizon, it is only required that 
the spacetime metric has the form of Eq. (\ref{eq:ds}) at some finite
region close to the horizon. This observation means that Eq. (\ref{eq:T3})
gives the temperature also for isolated horizons, i.e., for stationary
horizons in spacetimes which permit flux of gravitational radiation or
matter fields far from the horizons (for recent studies on isolated
horizons, see, for instance, Ashtekar et al in Ref. \cite{ash}). 

Besides the isolated horizons, a local view on Hawking radiation 
is necessary at least in the following two
situations. First of all, in spacetimes containing a black hole 
and a de Sitter horizon with unmatching surface gravities, it is
impossible to define an unambiguous temperature characterizing the 
whole spacetime. However, by treating both of the horizons separately, 
one can introduce local notions of temperature for these horizons 
through Eq. (\ref{eq:T3}). Note, though, that the spacetime in
question will not be in thermal equilibrium. Secondly, a local
approach to the Hawking radiation is needed in the presence of a naked
singularity. Such a situation arises, for instance, when one studies
the Hawking effect at the region inside the inner horizon of a
Reissner-Nordstr\"om black hole \cite{pm}. Again, an observer at 
rest close to the horizon will see a thermal flux of particles 
with the temperature (\ref{eq:T3}) coming out of the horizon. The
situation may become complicated, however, if one needs to describe 
the time evolution of the receding particles outside of the local
neighborhood of the horizon, because eventually 
one would also need to deal with the singularity. It is known from 
a work of Horowitz and Marolf that sometimes the boundary conditions 
of the Klein-Gordon field cannot be uniquely defined at the
singularity \cite{hm}. If this happens to be the case, there
is some loss of predictability in the theory, because it is not 
clear how the solutions representing particles at the horizon 
evolve at ``later times". Nevertheless, these ambiguities should 
not prevent us from defining the notions of a particle and thermal
radiation locally at the horizon, but indeed the ultimate fate
of the radiating particles may remain unspecified. 

Finally, let us comment on two possible generalizations of the
approach given in this paper. The form of Eq. (\ref{eq:ds}) does
not include certain important stationary spacetimes, such as the
Kerr-Newman spacetime. The idea of such generalization, however,
is probably quite easy to express. Since in the Kerr-Newman
spacetime the black hole horizon ``rotates" with a certain angular
velocity $\Omega_\text{H}$ about its symmetry axis, it is natural
to assume that the observers used in the analysis should rotate
along with the horizon, i.e., with the angular velocity $\Omega_\text{H}$
about the symmetry axis. The generalization should therefore be
straightforward but more laborious. One can also wonder what
happens if one of the functions $f(r)$ in Eq. (\ref{eq:ds}) is 
replaced by a different (but smooth) function $g(r)$, in order to
obtain more general form for the background metric. Curiously, for
certain class of functions $g(r)$ the answer can be found quite
easily \cite{gabor}. Consider a spacetime metric
\be \label{eq:fgmetric}ds^2 = -f(r)\, dt^2+\frac{dr^2}{g(r)}+dL^2_\perp .
\ee 
If the (smooth) functions $f(r)$ and $g(r)$ have the same 
root at $r=r_i$ and the ratio $f/g$ is always finite and positive,
one can define a new radial coordinate $\rho$ such that
\be \frac{dr^2}{g(r)} = \frac{d\rho^2}{f\big( r(\rho )\big)}.
\ee
With this coordinate transformation the spacetime metric becomes to
\be ds^2 = -\widetilde{f}(\rho )\, dt^2+\frac{d\rho^2}{\widetilde{f}(\rho )}+dL^2_\perp ,
\ee 
where we have denoted $\widetilde{f}(\rho ):=f\big( r(\rho )\big)$ and the
metric $dL^2_\perp$ can depend on $\rho$. When $dL^2_\perp = dx^2 + dy^2$,
the generalization is trivial: The situation is identical to the plane
symmetric case considered earlier. However, if $L^2_\perp = r^2\big( d\theta^2+\sin^2\theta\, d\varphi^2\big)$,
there will be a slight modification to the earlier calculations simply because now 
$r=r(\rho )$. It turns out, thought, that the only essential difference concerns
the potential analogous to that of Eq. (\ref{eq:tildeV}), which now behaves like
\be \widetilde{V}(\rho ) \xrightarrow{\rho \rightarrow \rho_i} \frac{\widetilde{f}'(\rho_i)\, r_i'}{r_i},
\ee
where $\rho_i$ is the root of $\widetilde{f}$ such that $r_i=r(\rho_i)$, 
the comma denotes a derivative with respect to the coordinate $\rho$, and
$r_i'=r'(\rho_i)$. When this potential is small enough to neglect, one
obtains the temperature
\be \label{eq:generalT}T = \frac{1}{\sqrt{\widetilde{f}(\rho)}}\frac{|\widetilde{f}'(\rho_i)|}{4\pi}
\ee
for the horizon. We note that this result is consistent with the findings of Ref.
\cite{hay}, where the local Hawking temperature has been derived using the Hamilton-Jacobi 
variant of the Parikh-Wilczek tunneling method. In fact, the results of that reference 
were obtained for general spherically symmetric spacetimes, holding also for 
non-static case, and when the spacetime metric is specified as in Eq. (\ref{eq:fgmetric}),
the results concur with the temperature (\ref{eq:generalT}). The fact that the Hawking 
temperature can be derived by the Hamilton-Jacobi method for a general spherically symmetric 
spacetime gives us a reason to believe that also the method used in this paper could be 
further generalized for arbitrary (and even time-depended) functions $f$ and $g$.

\begin{acknowledgments} The author would like to thank Gabor Kunstatter and Jarmo
 M\"akel\"a for constructive criticism during the preparation of this paper.
\end{acknowledgments}


\begin{thebibliography}{30}
\bibitem{obs} Similar idea has been already considered, in the case of the Schwarzschild 
	black hole, by Unruh in Ref. \cite{unruh}.
\bibitem{bif} This point is located at the bifurcation point of the future and the past horizon.
\end{thebibliography}
\end{document}